\documentclass{article}

\usepackage{PRIMEarxiv}

\usepackage[utf8]{inputenc} 
\usepackage[T1]{fontenc}    
\usepackage{hyperref}       
\usepackage{url}            
\usepackage{booktabs}       
\usepackage{amsfonts}       
\usepackage{nicefrac}       
\usepackage{microtype}      
\usepackage{lipsum}
\usepackage{fancyhdr}       
\usepackage{graphicx}       
\graphicspath{{media/}}     

\title{When Researchers Say Mental Model/Theory of Mind of AI, What Are They Really Talking About?
}

\author{
  Xiaoyun Yin, Elmira Zahmat Doost, Shiwen Zhou, Garima Arya Yadav, Jamie Gorman \\
  Center for Human, Artificial Intelligence, and Robot Teaming \\
  Arizona State Univeristy \\
  Gilbert\\
  \texttt{xiaoyun.yin@asu.edu} \\
}

\begin{document}
\maketitle

\begin{abstract}
When researchers claim AI systems possess ToM or mental models, they are fundamentally discussing behavioral predictions and bias corrections rather than genuine mental states. This position paper argues that the current discourse conflates sophisticated pattern matching with authentic cognition, missing a crucial distinction between simulation and experience. While recent studies show LLMs achieving human-level performance on ToM laboratory tasks, these results are based only on behavioral mimicry. More importantly, the entire testing paradigm may be flawed in applying individual human cognitive tests to AI systems, but assessing human cognition directly in the moment of human-AI interaction. I suggest shifting focus toward mutual ToM frameworks that acknowledge the simultaneous contributions of human cognition and AI algorithms, emphasizing the interaction dynamics, instead of testing AI in isolation.
\end{abstract}


Humans develop theories to explain each other's behaviors (Sellars, 1956). This ability to infer each other's mental states (Premack \& Woodruff, 1978) has been called the theory of mind (ToM). However, we are in direct continuous contact with our own minds but not the mental states of others; therefore, studying human mental states is difficult. This abstract concept is not related to any specific parameters or rubric; it's the theory of an experienced mental state that we can't observe in any specific event or attribute to different, distinct events.

Kosinski (2023) argues that individual artificial neurons in LLMs function like "Chinese rooms", which follow mathematical instructions without genuine understanding. The author also suggests that complex cognitive abilities may emerge at the network level, similar to how human cognition is assumed to emerge from networks of neurons. Mechanically speaking, this comparison makes sense. However, debating whether models are merely Chinese rooms may be unproductive. The key is to study how they interact, as humans typically don't question others' cognition during normal interaction. Meanwhile, Strachan et al. (2024) claim that LLM behavior is "indistinguishable from human behavior" on ToM tasks on various dimensions. However, if you listen to an authentic interaction between two humans and an interaction between a human and an AI agent, you can tell a difference. Under the assumption that a ToM is fundamental to human social interaction, rather than a byproduct of it, this finding has motivated researchers to develop benchmarks that offer a practical and cost-effective approach to assessing ToM capabilities in LLMs.

Researchers like Gu et al. (2024) tested GPT-4 using their SimpleToM dataset. While GPT-4 achieves approximately 90\% accuracy on ToM questions, performance drops to approximately 50\% on behavior prediction and 15\% on behavioral judgment. The test results are not ideal, and the test itself is also questionable, unless being able to answer a battery of ToM questions proves that GPT-4 has a ToM. The limitations of this kind of approaches has also been pointed out by Wang et al. (2025), that current ToM tasks have three major limitations: 1) Theoretically, ToM should be multidimensional but has only been tested in one dimension; 2) Current ToM tests lack construct validity; and 3) Evaluations always use third-person static scenarios rather than spontaneous dynamic interactions. They also report that 75.5\% of ToM measures focus only on beliefs rather than other considerations such as emotion, desires, and intentions. In a narrow technical sense, AI systems that track beliefs and predict behaviors could be said to have “ToM”. However, this technical capability differs fundamentally from human ToM grounded in embodied experience.

\section{The Fundamental Flaw in Cognitive Testing for LLMs }

The trend in evaluating LLMs using cognitive tasks may show a fundamental misunderstanding about what these systems actually are and what we need to know about them. Researchers designed ToM tests, reasoning challenges, and planning problems for LLMs by assuming that humans process information like LLMs do. But as Kambhampati (2024) argues, LLMs are fundamentally "n-gram models on steroids" performing "universal approximate retrieval" rather than reasoning focused on survival that humans do.

Consider how humans may develop ToM through embodied experience. Children learn that others have different perspectives through physical interactions such as hiding objects, playing peek-a-boo, and observing emotional responses. Psychologists study how children build up their abilities using developmental scales. For example, Wellman and Liu (2004) described scales of a sequence of tasks to test children's diverse desires, diverse beliefs, knowledge access, contents false belief, explicit false belief, belief-emotion, and real-apparent emotion. While this is representative work, it doesn't mean we should test LLMs in the same way or even compare them to humans (such as the SimToM prompting framework by Wilf et al. (2024)) by assuming these researchers are correct in reducing human interaction to their identified dimensions of a ToM.

Some researchers have designed benchmarks to measure ToM abilities specifically for LLMs, such as Kanishk et al. (2023) with their social reasoning benchmark (BigToM) and Chen et al. (2024) with their ToMBench. But why should we accept that LLMs even have a belief that can be tested? More importantly, what does passing these tests tell us about how AI will function in dynamic human-AI interactions?

Many benchmarks developed for testing LLMs' ToM are derived from psychological tests such as the Sally-Anne test (Scassellati, 2001) but lack validity and reliability. As ToM is proposed as an ability to understand and respond to ongoing feedback from the social environment, the evaluation of ToM should not be static (Wang et al., 2025). The tests are measuring something about ToM, certainly, but not meaningful interaction in a valid sense. They're measuring how well statistical patterns from human cognitive behavior in the training data can be reproduced in response to prompts. In other words, it is learning patterns that we intend it to, and we are asked to accept this as having a ToM that some researchers want it to.

This becomes clearer when we examine how LLMs fail. When planning problems are presented with obfuscated names, removing the statistical patterns from training data, LLM performance plummets dramatically. They're not reasoning inductively through the structure of problems; they're deductively matching surface patterns to similar examples in their training corpus. The same limitation appears in self-critique tasks. Despite claims that LLMs can verify their own reasoning, studies show performance actually worsens with self-verification as models “hallucinate” both false positives and false negatives.

The recent phenomenon that Cuadron et al. (2025) termed the "Reasoning-Action Dilemma" further exposes this limitation. Large Reasoning Models exhibit what appears to be "overthinking", where they bias internal reasoning over environmental feedback. But this isn't genuine overthinking in the human sense, such as in anxiety-driven rumination. It's the model getting stuck in loops of text generation that statistically resemble reasoning without genuine deliberation or decision-making authority.

The test is legitimate for measuring behavioral reproduction accuracy, but not valid for inferring genuine social cognitive processes or, more importantly, for predicting that AI functions as a human does in human contexts.

\section{Why Testing Misses the Point}

The distinction between simulation and authentic mental processes matters, but not for reasons typically discussed. LLMs are good at simulating ToM responses because they've been trained on billions of examples of human discourse about mental states aimed at generating those responses. They can produce outputs that seem to demonstrate understanding of beliefs, desires, and intentions. However, this cannot be called "ToM" in an adaptive human sense.

This distinction becomes clearer when examining motivated reasoning. Humans don't always reason correctly, even when they possess sufficient cognitive resources. We engage in “wishful thinking”, “confirmation bias”, and emotionally driven reasoning. As Kunda (1990) notes, human reasoning involves "hot motivation to reach desired goals" alongside "cold motivation favoring accuracy". AI systems, following optimization objectives, will lack these intrinsic motivational states. When AI systems make errors, it's not because desire overrides logic but because they've encountered edge cases that their training and algorithm do not cover.

Proponents of AI cognition often invoke “emergence”, which is a mysterious way of claiming that sufficiently complex neural networks spontaneously develop genuine understanding. They point to studies like Zhu et al. (2024) showing LLMs internally represent different agents' beliefs in their neural activations. However, finding structured representations doesn't prove consciousness exists any more than finding face-selective neurons proves the fusiform face area of the human brain experiences faces.

Debate about genuine versus simulated cognition distracts from the central issue: isolated ToM tests do not tell us what actually matters, nor how AI systems should function within human social and collaborative contexts.

\section{Mutual ToM}

Rather than asking whether AI "has" a ToM, we should examine how humans and AI systems can mutually develop understanding. Instead of validating AI performance on scales of ToM using isolated cognitive tests, we should test the system dynamics when AI operates within human environments. The critical question isn't whether AI can pass a false-belief task in isolation, but how human-AI interaction changes both behavior and outcomes when the human-AI system is working on a project rather than taking a test.

Wang and Goel's (2022) Mutual ToM framework offers one approach that preserves the ToM concept by focusing on three iteratively shaping elements: interpretation, feedback, and mutuality. In this framework, humans and AI agents construct representations of each other, but these representations serve different functions. Humans apply ToM to AI systems through anthropomorphism, attributing mental states that the AI doesn't possess. Meanwhile, AI systems build statistical models predicting human behavior without assuming any genuine understanding of mental states.

The key insight is that effective human-AI collaboration doesn't require AI to have a ToM but rather requires systems that support mutual adaptation and understanding. Zhang et al.'s (2024) findings support this approach. They found AI's independent ToM capability didn't significantly impact team performance. What mattered was enhancing human understanding of the agent. Surprisingly, bidirectional communication sometimes decreased performance, with participants reporting increased cognitive workload. This suggests that pretending AI has ToM might actually hinder rather than help human-AI collaboration. Considering the dynamic nature of mutual theory of mind, human-autonomy team (HAT) resilience studies employing dynamic measurements (e.g., Grimm et al., 2023) can serve as a valuable reference for future ToM studies, particularly those focusing on team-level states in human-AI collaboration. 

This distinction matters for practical purposes. We should stop asking "Does this AI understand false beliefs?" and start asking "How does this AI system change human behavior in collaborative settings?" Then stop endless philosophical debates about “emergent” consciousness and ToM test scores.

\section{Conclusion}

When researchers test AI’s ToM, they're observing sophisticated behavioral prediction, not genuine mental state attribution. But more importantly, they're asking the wrong question. The current approach of administering human cognitive tests to AI systems, whether adapted from child psychology or specially designed for LLMs, fundamentally misunderstands both what these systems are from a psychological perspective and what we need to know to use them effectively.

LLMs achieve remarkable performance on ToM tasks through statistical pattern matching across massive datasets, but this differs fundamentally from human experience, grounded in embodied experience, motivated reasoning, and genuine understanding, which led to the development of the ToM concept in the first place. However, even if AI does replicate ToM behaviors in testing conditions, these isolated assessments do not tell us about what actually matters: how human-AI systems function together as a cognitive system. Specifically, we should abandon attempts to improve AI's scores on ToM tests and instead study mutual adaptation and understanding between humans and AI systems. The mutual approach acknowledges both human tendencies to anthropomorphize and AI's statistical modeling capabilities, designing for effective collaboration within these constraints. The question isn't whether AI thinks like or simulates the human mind. It's how humans and AI can work together in the service of human survival and flourishing. After all, what matters isn't how well models score on ToM tests, but rather how good the models can serve us, humans.


\begin{thebibliography}{99}

\bibitem{Chen2024}
Chen, Z., Wu, J., Zhou, J., Wen, B., Bi, G., Jiang, G., Cao, Y., Hu, M., Lai, Y., Xiong, Z., \& Huang, M. (2024). 
ToMBench: Benchmarking theory of mind in large language models. 
\textit{arXiv preprint arXiv:2402.15052}.

\bibitem{Cuadron2025}
Cuadron, A., Li, D., Ma, W., Wang, X., Wang, Y., Zhuang, S., Liu, S., Schroeder, L. G., Xia, T., Mao, H., Thumiger, N., Desai, A., Stoica, I., Klimovic, A., Neubig, G., \& Gonzalez, J. E. (2025). 
The danger of overthinking: Examining the reasoning-action dilemma in agentic tasks. 
\textit{arXiv preprint arXiv:2502.08235}.

\bibitem{Gandhi2023}
Gandhi, K., Fränken, J. P., Gerstenberg, T., \& Goodman, N. (2023). 
Understanding social reasoning in language models with language models. 
In \textit{Advances in Neural Information Processing Systems} (Vol. 36, pp. 13518--13529).

\bibitem{Grimm2023}
Grimm, D. A., Gorman, J. C., Cooke, N. J., Demir, M., \& McNeese, N. J. (2023). 
Dynamical measurement of team resilience. 
\textit{Journal of Cognitive Engineering and Decision Making}, 17(4), 351--382. 

\bibitem{Gu2024}
Gu, Y., Tafjord, O., Kim, H., Moore, J., Bras, R. L., Clark, P., \& Choi, Y. (2024). 
SimpleTom: Exposing the gap between explicit tom inference and implicit tom application in LLMs. 
\textit{arXiv preprint arXiv:2410.13648}.

\bibitem{Kambhampati2024}
Kambhampati, S. (2024). 
Can large language models reason and plan? 
\textit{Annals of the New York Academy of Sciences}, 1534(1), 15--18.

\bibitem{Kosinski2023}
Kosinski, M. (2023). 
Theory of mind may have spontaneously emerged in large language models. 
\textit{arXiv preprint arXiv:2302.02083}.

\bibitem{Kunda1990}
Kunda, Z. (1990). 
The case for motivated reasoning. 
\textit{Psychological Bulletin}, 108(3), 480--498.

\bibitem{Premack1978}
Premack, D., \& Woodruff, G. (1978). 
Does the chimpanzee have a theory of mind? 
\textit{Behavioral and Brain Sciences}, 1(4), 515--526.

\bibitem{Scassellati2001}
Scassellati, B. M. (2001). 
\textit{Foundations for a Theory of Mind for a Humanoid Robot} (Doctoral dissertation, Massachusetts Institute of Technology).

\bibitem{Sellars1956}
Sellars, W. (1956). 
Empiricism and the philosophy of mind. 
\textit{Minnesota Studies in the Philosophy of Science}, 1(19), 253--329.

\bibitem{Strachan2024}
Strachan, J. W. A., Albergo, D., Borghini, G., Pansardi, O., Scaliti, E., Gupta, S., Saxena, K., Rufo, A., Panzeri, S., Manzi, G., Graziano, M. S. A., \& Becchio, C. (2024). 
Testing theory of mind in large language models and humans. 
\textit{Nature Human Behaviour}, 8(7), 1285--1295.

\bibitem{WangGoel2022}
Wang, Q., \& Goel, A. K. (2022). 
Mutual theory of mind for human-AI communication. 
\textit{arXiv preprint arXiv:2210.03842}.

\bibitem{Wang2025}
Wang, Q., Zhou, X., Sap, M., Forlizzi, J., \& Shen, H. (2025). 
Rethinking theory of mind benchmarks for LLMs: Towards a user-centered perspective. 
\textit{arXiv preprint arXiv:2504.10839}.

\bibitem{Wellman2004}
Wellman, H. M., \& Liu, D. (2004). 
Scaling of theory-of-mind tasks. 
\textit{Child Development}, 75(2), 523--541.

\bibitem{Wilf2023}
Wilf, A., Lee, S. S., Liang, P. P., \& Morency, L. P. (2023). 
Think twice: Perspective-taking improves large language models' theory-of-mind capabilities. 
\textit{arXiv preprint arXiv:2311.10227}.

\bibitem{Zhang2024}
Zhang, S., Wang, X., Zhang, W., Chen, Y., Gao, L., Wang, D., Zhang, W., Wang, X., \& Wen, Y. (2024). 
Mutual theory of mind in human-AI collaboration: An empirical study with LLM-driven AI agents in a real-time shared workspace task. 
\textit{arXiv preprint arXiv:2409.08811}. 

\bibitem{Zhu2024}
Zhu, W., Zhang, Z., \& Wang, Y. (2024). 
Language models represent beliefs of self and others. 
\textit{arXiv preprint arXiv:2402.18496}.

\end{thebibliography}
\end{document}